## Title

# Frequency- and time-resolved second order quantum coherence function of IDTBT single-molecule fluorescence

## Authors

Quanwei Li[1,2], Yuping Shi[3,4], Lam Lam[2,4,5], K. Birgitta Whaley[1,2]*, and Graham Fleming[1,2,4]*

## Affiliations

[1] Department of Chemistry, University of California, Berkeley, CA 94720, USA

[2] Kavli Energy Nanoscience Institute at Berkeley, Berkeley, CA 94720, USA

[3] Max Planck Institute for Polymer Research, Mainz 55128, Germany

[4] Molecular Biophysics and Integrated Bioimaging Division, Lawrence Berkeley National Laboratory, Berkeley, CA 94720, USA

[5] Graduate Group in Biophysics, University of California, Berkeley, CA 94720, USA

[*] Corresponding author. Email: whaley@berkeley.edu, grfleming@lbl.gov

**Abstract**

The frequency- and time-resolved second order quantum coherence function ($g^{(2)}(\tau)$) of single-molecule fluorescence has recently been proposed as a powerful new quantum light spectroscopy that can reveal intrinsic quantum coherence in excitation energy transfer in molecular systems ranging from simple dimers to photosynthetic complexes. Yet, no experiments have been reported to date. Here, we have developed a single-molecule fluorescence $g^{(2)}(\tau)$ quantum light spectroscopy (SMFg2-QLS) that can simultaneously measure the fluorescence intensity, lifetime, spectra, and $g^{(2)}(\tau)$ with frequency resolution, for a single molecule in a controlled environment at both room temperature and cryogenic temperature. As a proof of principle, we have studied single molecules of IDTBT (indacenodithiophene-co-benzothiadiazole), a semiconducting donor-acceptor conjugated copolymer with a chain-like structure that shows a high carrier mobility and annihilation-limited long-range exciton transport. We have observed different $g^{(2)}(\tau=0)$ values with different bands or bandwidths of the single molecule IDTBT fluorescence. The general features are consistent with theoretical predictions and suggest non-trivial excited state quantum dynamics, possibly showing quantum coherence, although further analysis and confirmation will require additional theoretical calculations that take into account the complexity and inhomogeneity of individual IDTBT single molecular chains. Our results demonstrate the feasibility and promise of frequency- and time-resolved SMFg2-QLS to provide new insights into molecular quantum dynamics and to reveal signatures of intrinsic quantum coherence in photosynthetic light harvesting that are independent of the nature of the light excitation.

## Introduction

Questions about the roles of quantum effects in complex systems such as large molecules and living organisms have fascinated scientists for nearly a century[1–3]. Quantum effects are usually observed under extreme conditions where closed single atomic systems or solid crystals are well isolated from the environment (e.g., in vacuum) and often cooled to cryogenic temperatures to remove thermal noise. In stark contrast, complex systems involving large molecules and/or biological processes are often found in complex open environments with rapid decoherence induced by phonons or vibrations, especially at room temperature. Thus, properties and dynamics of complex systems can often be modeled by classical approaches without invoking the underlying quantum mechanics that would arise at the atomic level. It has been long sought to discover experimental evidence for *explicit* or *non-trivial* quantum effects such as coherence and entanglement in complex systems that *can only* be understood by quantum mechanics, with no adequate classical models. The observations of long-lived oscillations in ultrafast spectroscopy over 15 years ago suggested that quantum coherence occurs in the excitation energy transfer in photosynthetic complexes[4–6], triggering renewed interest in the manifestation and role of quantum effects in biological systems[7]. Further studies since then have led to the emerging consensus that the observed quantum coherence in photosynthetic systems largely reflects vibronic coherence[8–12]. However, it is very difficult using laser-based ultrafast nonlinear spectroscopies to unambiguously distinguish *intrinsic* quantum coherence from laser induced *extrinsic* coherent dynamics and associated artifacts.

Using the coherence functions of emitted light to reveal features of the excited state dynamics in matter is a longstanding goal[13-24] that has recently attracted theoretical attention within the photosynthetic community. Recent proposals suggest that a quantum light spectroscopy measuring frequency- and/or polarization- resolved single-molecule fluorescence photon statistics could reveal molecular quantum coherence, with model examples presented under excitation by either lasers[25] or weak thermal sources[26-28]. These predictions that such measurements can contain similar dynamic information about a complex system as ultrafast nonlinear spectroscopies are very attractive, since measuring coherence functions of the emitted light does not require a coherent source to excite the system and thus such experiments possess the advantage of removing any possible laser-induced coherent dynamics and potential associated artifacts. In particular, Olaya-Castro and coworkers[27] have analyzed the frequency-resolved fluorescence $g^{(2)}(\tau)$ for a prototypical vibronic heterodimer model, finding that vibronic quantum coherence can break the microscopic detailed balance and manifest itself in an asymmetric shape of the frequency-resolved $g^{(2)}(\omega_1, \omega_2, \tau)$ on sub ps timescales which can thereby serve as a witness for a coherent vibronic contribution to electronic excitation transport. Importantly, they also found that $g^{(2)}(\tau = 0)$ (over the time scale of fluorescence lifetime ~ ns) has a large and sensitive dependence on (1) the coupling between the exciton and the vibrational mode, (2) the resonance between the vibrational mode energy and the energy difference between two excitons, (3) the exciton delocalization, and (4) the filtered detection frequency. Another proposed signature of excitonic or vibronic coherence is the time asymmetry in a polarization-resolved $g^{(2)}(\tau)$ measurement of single-molecule fluorescence, with calculations for a realistic model of FMO with non-Markovian coupling to phonons showing such asymmetry within a few ps near $\tau = 0$ [28]. Taken together, these theoretical results show that $g^{(2)}(\tau = 0)$ (~ ns time scale) may already witness coherent coupling.

Such proposed single-molecule experiments are quite distinct from our previous single-photon experiments with a molecular ensemble[29-30], requiring single molecule capabilities together with fast single photon detection and time correlation capabilities, rather than heralded single photon sources. Here, we report the development of such a single molecule $g^{(2)}(\tau)$ (SMFg2) setup, and the measurement of the frequency- and time-resolved $g^{(2)}(\tau)$ of fluorescence from single molecules of IDTBT semiconducting polymer on a ns time scale at both room temperature and cryogenic temperature. This first experiment is designed to establish the technique on a model system having high optical efficiency and using standard photodiode detectors, before moving to the ps time scale, which requires higher quality detectors.

The model system IDTBT (indacenodithiophene-co-benzothiadiazole), is a semiconducting donor-acceptor conjugated copolymer with a chain-like structure that shows a high carrier mobility and annihilation-limited long-range exciton transport,[31] and is a promising material for photovoltaic and optoelectronics[32], biosensing[33-34], and even biophotocatalysis[35]. Using weak pulsed laser excitation, we have observed different $g^{(2)}(\tau = 0)$ values in different regions of the emission spectrum or with different detection bandwidths of the single molecule fluorescence. Such general features are consistent with the theoretical predictions of Ref. 15 and might imply non-trivial quantum dynamics in the excited state dynamics. At cryogenic temperatures (~100K) some molecules show significantly narrowed emission spectra and have significantly smaller $g^{(2)}(\tau = 0)$ values. However, we also find that the specific $g^{(2)}(\tau = 0)$ values vary for different single molecules at both room and cryogenic temperatures. Such individual molecular variations are not unexpected, given the complexity and inhomogeneity of individual IDTBT single molecular chains. We analyze the results here in terms of numbers of emitters on the single molecular chain, showing that the $g^{(2)}(\tau = 0)$ measurements can distinguish between single or multiple emitters on a single IDBT molecular chain and can also reveal low temperature changes in numbers of emitters consistent with exchange narrowing.

## Results

### The experimental setup

Combining and extending single-molecule spectroscopy[36-37] and quantum optics, we developed a SMFg2 setup that can perform synchronized measurement of the following single-molecule fluorescence properties: (1) intensity, (2) lifetime, (3) spectra, and (4) second order quantum coherence function $g^{(2)}(\tau)$, with frequency and/or polarization resolution. The experimental setup functions in a controlled environment that can be held either at room temperature or at cryogenic temperature (Fig. 1(a)). Given the heterogeneity of the distribution of single molecules and the temporal fluctuation of this distribution as well as fluctuations of individual molecules, it is essential to measure all of these observables simultaneously. The setup is based on a home-built multi-functional scanning confocal microscope (see Methods). It employs an oil inversion high numerical aperture objective for high spatial resolution and high fluorescence collection efficiency. Spatial scanning and sample positioning is controlled by a piezo stage. The setup has three independent, simultaneous outputs. One output is connected to a spectrometer for emission spectrum measurements. Each of the other two outputs is connected to a single-photon detector (~ 350 ps time resolution) and we can insert a set of spectral filters and/or a polarizer to enable frequency and/or polarization resolution of each of these signals.

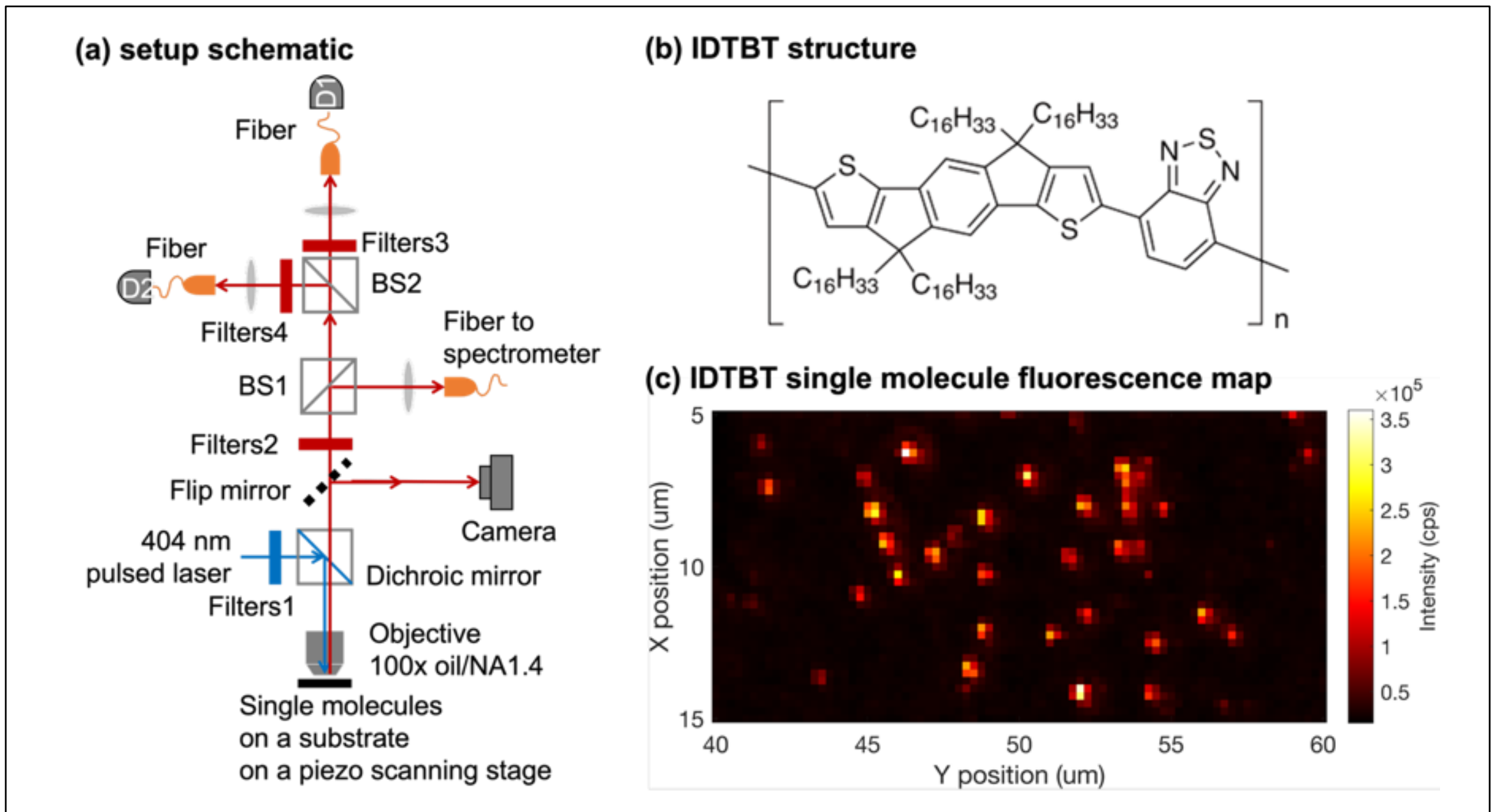

**Fig. 1.** Principle of the experiment to measure frequency- and/or polarization resolved $g^{(2)}(\tau)$ for single molecules at room or cryogenic temperatures. (a) Simplified schematic of the experimental setup. BS denotes beamsplitter. D1 and D2 are the two detectors used for synchronized TCSPC measurements. (b) The molecular structure of the IDTBT polymer sample used in our experiments. The average chain length is n~110 monomers (Methods). (c) A typical IDTBT fluorescence spatial map showing individual single molecules.

Fig. 1(b) shows the molecular structure of the studied molecular system IDTBT. We have recently studied long range excitonic transport in thin films of IDTBT, probing the role of exciton annihilation dynamics[31]. For the SMFg2 setup we adopted established methods in single-molecule spectroscopy and imaging to immobilize single molecule[36-37]. The single molecule samples were prepared in a glove box that excludes oxygen and water. The blend of pico-molar IDTBT solution and polymethyl methacrylate (PMMA) solution is spin-coated on a cover glass and then annealed to leave single molecules of IDTBT embed in the supporting solid matrix of PMMA. The average chain length of the IDTBT molecules is ~110 (Methods). The sample is then encapsulated and sealed by epoxy before being taken out of the glove box for measurement. See Methods for details of the sample preparation.

Fig. 1(c) shows a typical spatial intensity map of the IDTBT single molecule fluorescence. Many but not all of the isolated emission spots contain a single IDTBT molecule, which is confirmed by measuring their emission spectrum. Due to its excellent optical properties[38], a single molecule of IDTBT is very bright and has a high detected fluorescence intensity of the order of $10^5$ counts per second. However, an aggregate of molecules has a significantly higher fluorescence intensity, allowing the single molecules to be distinguished by their fluorescent count rate. To measure a specific single molecule, it is moved by the piezo stage to the fluorescence excitation and collection spot. Our analysis below focuses on the single molecule spectra.

IDTBT single molecule fluorescence spectra, intensity, and lifetime

After spatially isolating individual IDTBT single molecules, we characterized the time traces of their individual fluorescence spectra, intensity, and lifetime in a synchronized measurement (Fig. 2). All of the three IDTBT single molecules presented in Fig. 2 have a similar fluorescence spectrum that shows a dominant peak near 700 nm which is attributed to the 0-0 vibronic transition of the $S_0 \rightarrow S_1$ electronic transition, and a small shoulder near 760 nm which is attributed to the corresponding 0-1 vibronic transition (Fig. 2(a), (d), (g)), consistent with earlier spectral results of diluted IDTBT solution (see SI of Ref. 31). Blinking of individual single molecules is clearly visible within the time trace of the fluorescence spectrum, manifesting itself as random dimming and brightening of the fluorescence emission[39-41]. Note that the blinking effect mainly changes the overall intensity of the fluorescence rather than its spectral shape (i.e., the width in wavelength of panels (a), (d), (g) in Fig. 2). This suggests that blinking is mainly caused by the turning-on of a nearby trap state or the proximity of a quencher molecule, rather than by a switch or change of the emissive state itself.

Panels (b), (e), (h) in Fig. 2 shows the fluorescence intensity obtained from synchronized time-correlated single-photon counting (TCSPC) measurement with 2 single photon detectors. These measurements clearly show that both detection channels follow the same intensity fluctuations that are seen in the spectrum time trace data, as expected for direct decay to the ground state. We also obtain the fluorescence lifetime in each time interval of 1 s (panels (c), (f), (i) in Fig. 2) by fitting the TCSPC data within that time interval to the convolution of a single exponential decay and a Gaussian instrument response function (see Methods). The fluorescence lifetime of each individual single molecule is somewhat different, suggesting variable individual total decay rates that are likely due to the heterogeneity of the single molecule distribution. The measured lifetimes range from 0.37 ns to 1.35 ns. Fits to the smallest and largest fluorescence lifetime for each molecule over the 10 s measurement time are reported in Table S1 of the Supplementary

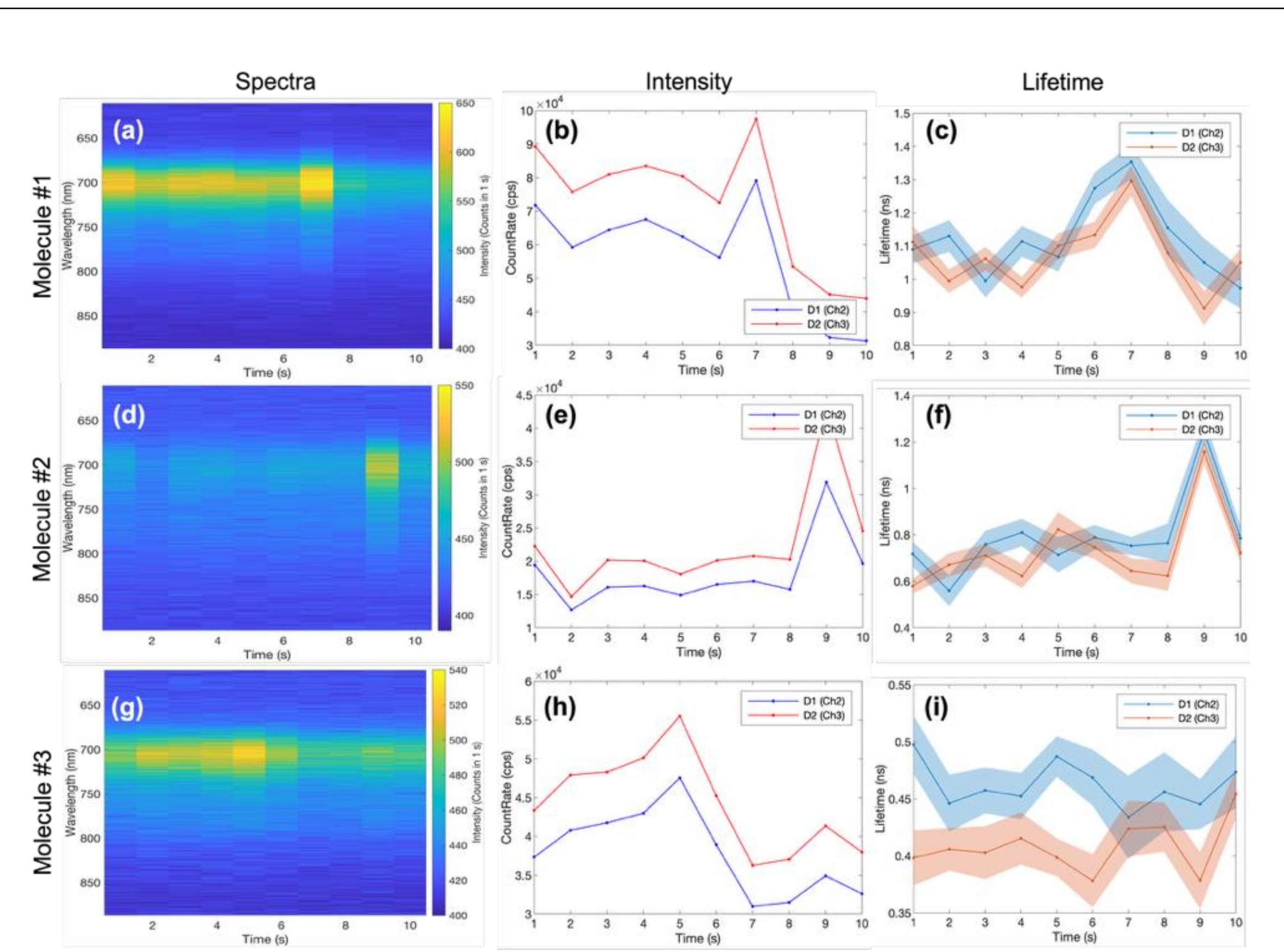

**Fig. 2. Single molecule fluorescence spectra for 3 individual molecules of IDTBT.** Panels (a, d, g): time trace of single molecule fluorescence spectra on a time scale of s. Panels (b, e, h): intensity of emission as a function of time in s. Panels (c, f, i): lifetime (ns) as function of time in s, showing blinking over timescales longer than the lifetime.

Material and fits extracting the fluorescence lifetimes for all time intervals are reported in Table S2. The variable ns fluorescence lifetimes of single molecules #1 and #2 also appear to closely follow the trend of their fluorescence intensity over the s timescale,
consistent with the blinking not being caused by the change of the emissive state itself. In contrast, the fluorescence lifetime of single molecule #3 appears to be approximately constant over the s timescale, regardless of its blinking. This appears to suggest that for this molecule, the emissive state has changed to some extent in the time period between 5 s and 7 s. This is also consistent with the disappearance of the narrow peak in the fluorescence spectrum of this molecule that is evident in Fig. 2(g).

IDTBT single molecule fluorescence $g^{(2)}(\tau)$ measured in different emission bands or with differing detection bandwidths

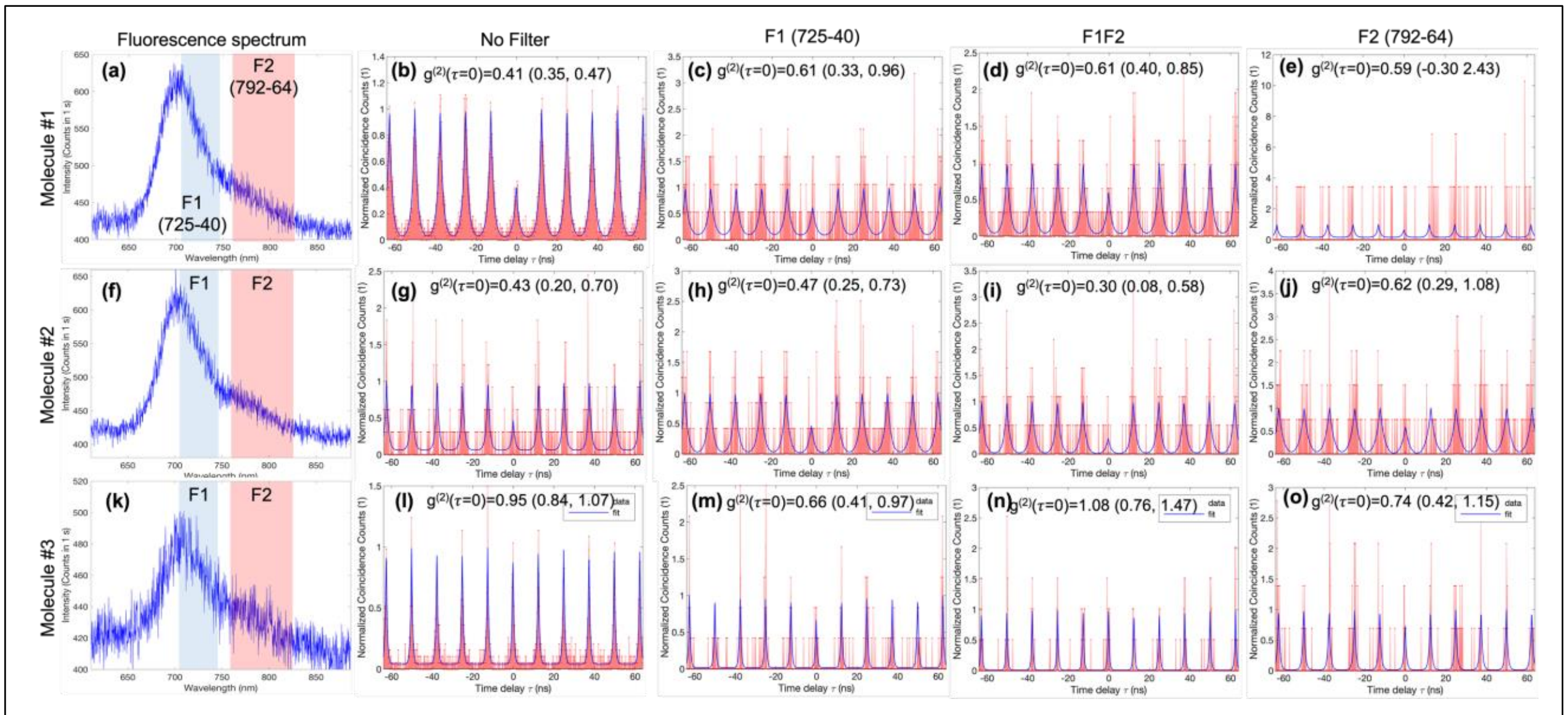

**Fig. 3.** IDTBT single molecule fluorescence $g^{(2)}(\tau)$ measured with different frequency bands or bandwidths. Panels (a, f, k): fluorescence spectra of the 3 single molecules from Fig. 2, with superposed transmission bands of the two bandpass filters F1 (40 nm bandwidth centered at 725 nm, blue shadow) and F2 (64 nm bandwidth centered at 792 nm, red shadow). Panels (b, g, l): $g^{(2)}(\tau)$ measured without any bandpass filters. Panels (c, h, m): $g^{(2)}(\tau)$ with filter F1 located before both detectors. Panels (d, i, n): $g^{(2)}(\tau)$ measured with F1 located before detector 1 and F2 located before detector 2. Panels (e, j, o): $g^{(2)}(\tau)$ measured with F2 located before both detectors. For all $g^{(2)}(\tau)$ plots, the red dots are measured data with time bin 128 ps, the red shadow is the area under the dots, and the blue curves represent fitting of the data to equally spaced double-side single exponential peaks as described in Methods. The values of $g^{(2)}(\tau = 0)$ extracted from these fits are shown at the top of each plot, with the values in parentheses representing 95% confidence intervals.

We then measure the room-temperature frequency- and time-resolved $g^{(2)}(\tau)$ of the same 3 single molecules that are presented in Fig. 2. The measured typical fluorescence spectra of these 3 single molecules are shown in panels (a, f, k) of Fig. 3, respectively. The transmission bands of the two bandpass filters used to select the 0-0 (F1) and 0-1 (F2) transitions, respectively, are marked on each spectrum. For each single molecule, we have measured 4 different combinations of the selected frequency for the two detectors. These are as follows: (1) without any bandpass filter before any detector; (2) inserting the 0-0 (F1) bandpass filter before both detectors; (3) inserting the 0-0 (F1) bandpass filter before detector 1 and the 0-1 (F2) bandpass filter before detector 2; and (4) inserting the 0-1 (F2) bandpass filter before both detectors. The measured data were fitted by 11 equally spaced double-sided single exponential peaks with constant background (see Methods), yielding the blue curves from which the $g^{(2)}(\tau = 0)$ values reported in the figures were extracted. Without any bandpass filters, the $g^{(2)}(\tau = 0)$ values of single molecules #1 and #2 are well below the single-photon threshold 0.5, consistent with single molecules #1 and #2 being single emitters[42-44], while $g^{(2)}(\tau = 0)$ value of single molecule #3 appears to be close to 1, consistent with containing multiple emitters.

When frequency resolution is added, the $g^{(2)}(\tau)$ data now reveal an interesting generic behavior, namely that different values of $g^{(2)}(\tau = 0)$ are obtained when using fluorescence filters to detect different emission bands, as well when using different bandwidths for detection. The frequency-filtered of $g^{(2)}(\omega_1, \omega_2, \tau)$ data show that the zero-time values $g^{(2)}(\omega_1, \omega_2, 0)$ are consistent with the weakly anti-bunched values obtained in the calculations of the Olaya-Castro group for a vibronic dimer with intermediate values of the exciton-vibration coupling (see Fig. 4c in Ref. 27). However, the specific $g^{(2)}(\tau = 0)$ values are seen to be quite different for different single molecules. We attribute this to the complexity and heterogeneity of the single molecules of IDTBT, with the latter resulting from variable chain length, conformational differences, and variable local environmental factors such as charge, strain, and quenchers, as well as to the fact that the single molecules of IDTBT are immobilized in a solid matrix of PMMA.

IDTBT single molecule fluorescence $g^{(2)}(\tau)$ at cryogenic temperature

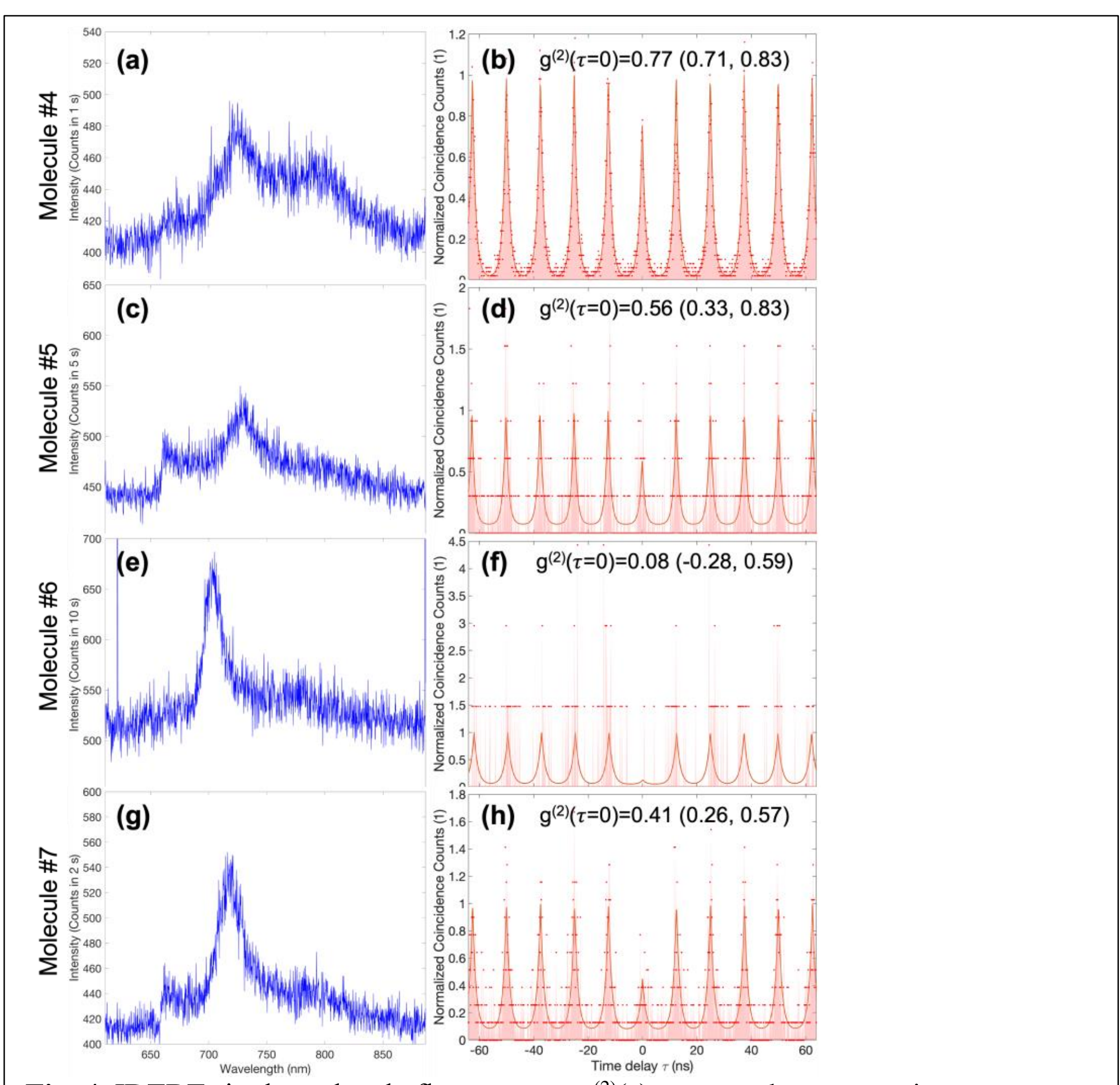

**Fig. 4.** IDTBT single molecule fluorescence $g^{(2)}(\tau)$ measured at cryogenic temperature. Panels (a, c, e, g): fluorescence spectra of 4 single molecules. Panels (b, d, f, h): broad band $g^{(2)}(\tau)$ measured without any bandpass filters. For the $g^{(2)}(\tau)$ plots, the red dots are measured data with time bin 128 ps and the red shadow is the area under the dots. The red curves represent fitting of the data to equally spaced double-side single exponential peaks as described in Methods and yield the extracted $g^{(2)}(\tau = 0)$ values shown at the top of each plot, with the values in parentheses representing 95% confidence intervals.

We further measured IDTBT single molecule fluorescence $g^{(2)}(\tau)$ at a cryogenic temperature of ~100 K (Fig. 4). For cryogenic measurement, the sample is mounted inside a cryostat, and a long working distance microscope objective is mounted on the piezo stage for scanning and position. The cryostat is cooled by a constant flow of liquid nitrogen and the temperature at the sample position is estimated to be about ~ 100 K. The change in setup results in much less collected fluorescence signal compared with our room temperature measurement and under these conditions we have only measured the broad band $g^{(2)}(\tau)$ without any bandpass filters.

Among the 4 typical single molecules measured at this temperature, Molecules #4 and #5 clearly show 2 fluorescence peaks, while their $g^{(2)}(\tau = 0)$ values lie between 0.5 – 1. This is consistent with multiple emitters or emissive states, emitting at different frequencies. Interestingly, Molecules #6 and #7 essentially show only one peak in their fluorescence spectra, which also has a much narrower line width compared with that of the typical room temperature spectra shown in Fig 3. These two molecules also show significantly lower $g^{(2)}(\tau = 0)$ values compared with the room temperature results, consistent with single independent emitters. The fluorescence narrowing - often referred to as 'exchange narrowing' - is known to result from the suppression of dynamic disorder at lower temperatures resulting in dominance of the intrinsic delocalization resulting from excitonic exchange and merging of nearby emitters.[45-46] This suggests that single molecules of IDTBT may show a temperature dependent change from multiple emitters at room temperature to single emitters at cryogenic temperatures. We note that some molecules do still have $g^{(2)}(0) \sim 1$, and hence multiple emitters, at ~ 100K. Such molecules may have kinks or sharp conformational changes that hinder delocalization.

**Conclusion and perspectives**

We have developed a SMFg2 spectroscopy that can simultaneously measure the intensity, lifetime, spectra, and $g^{(2)}(\tau)$ with frequency resolution of a single molecule in a controlled environment at either room temperature or cryogenic temperature. We have used this to study single molecules of the semiconductor donor-acceptor conjugated copolymer IDTBT with weak pulsed laser excitation. Our method enables detection and differentiation of both single molecule single photon emitters as well as single molecules that are two or more photon emitters. Our wavelength-resolved $g^{(2)}(\tau = 0)$ values are consistent with recent theoretical predictions[27] and indicate the potential of the frequency-resolved second-order correlation function $g^{(2)}(\omega_1, \omega_2, \tau)$ at variable time delay $\tau$ as a new tool to study complex molecular dynamics. Such measurements will need to be paired with sophisticated theoretical calculations of the second order fluorescence correlation function that take into account the complexity and inhomogeneity of an IDTBT single molecule chain, including exciton-phonon couplings. If it becomes possible to measure $g^{(2)}(\tau)$ with < 1 ps detection time resolution, this will allow studies of, for example, whether photosynthetic light harvesting makes use of quantum coherence, independently of the nature of the excitation, i.e., independent of whether the system is coherently excited by laser light or incoherently excited by a thermal source. Our results demonstrate the feasibility and promise of frequency- and time-resolved single molecule fluorescence $g^{(2)}(\tau)$ to provide new insights into molecular excited state quantum dynamics and to pave the way toward a new spectroscopy of complex coupled molecular systems.

This work was supported by the US Department of Energy, Office of Science, Basic Energy Sciences, Chemical Sciences, Geosciences and Biosciences Division, Solar Photochemistry Program (FWP 449A) and Photosynthetic Systems Program (award DE-SC001--9728).

## AUTHOR DECLARATIONS

### Conflict of Interest

The authors have no conflicts to disclose.

### Author Contributions

Q.L., G.R.F., and K.B.W. conceived the project. Q.L. designed and built the setup, performed the measurements, and analyzed the data. Y.S., L.L., and Q.L. prepared the sample. Q.L., G.R.F., K.B.W., and Y.S. discussed the results. Q.L., G.R.F., and K.B.W. wrote the manuscript with inputs from all other authors. All authors commented on the manuscript. G.R.F. and K.B.W. supervised the project.

### Corresponding authors

Correspondence and requests for materials should be addressed to Graham R. Fleming or K. Birgitta Whaley.

## DATA AVAILABILITY

The data that support the findings of this study are available from the corresponding authors upon reasonable request.

## Methods

Sample Preparation

The encapsulated samples of isolated IDTBT single chains in a poly (methyl methacrylate) (PMMA) matrix were fabricated under a nitrogen atmosphere in a well-operated glovebox. The IDTBT solution sample was prepared by dissolving IDTBT powder ($M_w$ = 346,000, polydispersity index, PDI = 2.43) in anhydrous Tetrahydrofuran (THF) and stirring at 60 °C overnight to form a clear 1.0 mg/mL solution. This gives polymeric chains with an average of 110 monomers (obtained from the number average molecular weight $M_n = M_w/PDI$ = 142,000 and dividing this by the molecular weight of the monomer). The resulting IDTBT solution was diluted 1,000,000 times with anhydrous THF to achieve a final concentration of 1.0 ng/mL, which was well-mixed with 50× volume of a syringe-filtered 50 mg/mL PMMA solution in anhydrous THF, yielding a clear IDTBT/PMMA blend solution for the subsequent spin coating fabrication (2000 rpm for 1 min) of a blend film on a pre-cleaned cover glass. The spin-coated IDTBT/PMMA blend films were annealed first at 70 °C for 15 min and then at 110 °C for 30 min, before implementing sample encapsulation with a pre-cleaned slide glass and UV-curable glue dispersed into a continuous glue line near the edge of the cover glass. During the UV-curing process, the LED UV (365 nm) irradiation was completely blocked across a large central area in the IDTBT/PMMA blend films to prevent photo-degradation of IDTBT single molecules.

The experimental setup

The setup is a home-built multi-functional scanning confocal microscope. The 80 MHz 808 nm fs laser pulses (Vitara, Coherent) are upconverted into 404 nm through second harmonic generation in a BBO crystal. The 404 nm laser is spatially and spectrally separated from the 808 nm residue and coupled into a single mode fiber to be used as the excitation source. Additional filters are used to further clean the laser (Filters1 = FF01-424/SP-25 and FF01-400/40-25, Semrock). A dichroic mirror (DMLP505R, Thorlabs) reflects the 404 nm laser through an oil immersion objective (100x oil/numerical aperture 1.4) onto the single molecule sample. The power of 404 nm laser is about 0.4 $\mu$W at the sample position with an estimated focus spot size about 300 nm. Sample spatial scanning and positioning is controlled by a piezo stage (P-611.3SF, PI). Fluorescence emission is collected by the same objective and transmitted through the dichroic mirror. A flip mirror and a EMCCD (Luca R, Andor) are used to image the laser residue and fluorescence to ensure good alignment. Two long pass filters (Filters2 = 490LP and LP610-25.4, MIDOPT) are used to block residual laser light before the florescence signal is split by beamsplitter 1 and directed one output through a 100-um core multimode fiber into a spectrometer (SP2300i + PIXIS400BR, Princeton Instrument). The fluorescence from the other output of beamsplitter 1 is further split by beamsplitter 2 and another filter (Filters3 or Filters4 = FF01-725/40-25 or FF01-792/64-25, Semrock) may be added before the fluorescence is coupled into a multimode fiber into detector 1 (50 um core fiber, SPCM AQRH 16, Excelitas) and detector 2 (100 um core fiber, SPCM AQRH 16, Excelitas). Each detection event from the two detectors is recorded with ~350 ps resolution by a time tagger (Time Tagger Ultra, Performance Edition, Swabian Instruments) with channel dead time setting 86 ns for lifetime and correlation measurements.

Data analysis

The fluorescence lifetimes reported in Fig. 2 were fitted by convolution of the raw data fluorescence decay histograms between single exponential and Gaussian functions) with 5 free parameters:

(A/2)*exp(c1^2/(4*tau^2))*exp(-(x - b1)/tau)*(1+erf((x -b1 - c1^2/(2*tau))/c1)) + d1,

where "A" is the overall height, "b1" is the peak position, "tau" is the decay lifetime, "c1" represents the Gaussian width, and "d1" is a constant background, which directly produced the lifetime values in tau. The resulting fluorescent lifetimes and fit parameters for the smallest and largest measured lifetimes for each molecule in Fig. 2 are reported in Table S1 of the Supplementary Material and the lifetimes and fit parameters of the full data set measured over 10 s are reported in Table S2.

For the second order quantum coherence functions $g^{(2)}(\tau)$ presented in Figs. 3 and 4, the raw coincidence count data were each fit by 11 equally spaced double-sided single exponential peaks with 6 free parameters:

a*exp(-b*abs(x+f))
+d*exp(-b*abs(x+f-e))+d*exp(-b*abs(x+f+e))
+d*exp(-b*abs(x+f-2*e))+d*exp(-b*abs(x+f+2*e))

+d*exp(-b*abs(x+f-3*e))+d*exp(-b*abs(x+f+3*e))

+d*exp(-b*abs(x+f-4*e))+d*exp(-b*abs(x+f+4*e))

+d*exp(-b*abs(x+f-5*e))+d*exp(-b*abs(x+f+5*e))

+c,

where “a” is the height of the center peak, “1/b” is the single exponential lifetime, “c” is the constant background, “d” is the equal height of all side peaks, “e” is the period between peaks, and “f” is the shift of center peak. The $g^{(2)}(\tau)$ data shown in Figs. 3 and 4 were normalized by the fitted equal height parameter “d” of all side peaks. The value of $g^{(2)}(\tau=0)$ is extracted from the ratio of the central peak height to the side peak height, “a/d”. Values of $g^{(2)}(\tau=0)$ and the fitting parameters for data in Fig. 3 and Fig. 4 are given in Tables S3 and S4 of the Supplementary Material.